# The response of disordered matter to electromagnetic fields

P. Lunkenheimer and A. Loidl
*Experimentalphysik V, Universität Augsburg, D-86135 Augsburg, Germany*

We have studied a variety of different disordered materials, including molecular and ionic liquids, supercooled liquids and glasses, ionic conductors, and doped semiconductors, in ac electromagnetic fields over an exceptional broad dynamic range including the rarely investigated GHz to THz region. We show that all classes of disordered matter exhibit an astonishingly similar response: In addition to Jonscher´s time-honored "Universal Dielectric Response", a superlinear power-law increase of the frequency-dependent conductivity shows up bridging the gap between the classical dielectric and the infrared region. We demonstrate that the universal dielectric behavior of disordered matter extends up to much higher frequencies than thought until now and covers the range from ultra-low (mHz) to phonon (THz) frequencies.

PACS numbers: 66.30.Hs, 72.20.Ee, 72.80.Ng, 77.22.Gm

Disordered matter is ubiquitous in our daily live. Especially its electrical applications are numerous, e.g. the use of doped crystalline and amorphous semiconductors as standard materials for electronics or solar cells, glasses and polymers as insulator materials, or ionic conductors as solid electrolytes for accumulators and fuel cells. As an increasing number of devices (for example in wireless communication systems or computer techniques) operate at progressively higher frequencies, the knowledge of the response of disordered matter to electromagnetic ac fields, in an as broad frequency range as possible is of high technical relevance. But the ac-response of disordered matter is also interesting from a theoretical point of view, especially in light of the fact that in the past years a number of universalities were found, indicating common microscopic processes in such different materials as, e.g., supercooled liquids, and doped semiconductors [1-5]. However, so far these universalities are restricted to certain classes of materials and/or limited frequency regimes, usually not extending beyond the GHz regime. Especially there is very few information about the increasingly important, however experimentally difficult to access frequency region between the range of classical dielectric spectroscopy, up to MHz-GHz frequencies, and the infrared region, accessible by optical experiments. In a time-honored review article, Jonscher [1] demonstrated that the response of disordered condensed matter to ac electromagnetic fields in the sub-GHz regime follows the so-called "universal dielectric response" (UDR), showing up, e.g., as a sublinear power law in the frequency-dependent conductivity. In some cases and for special classes of materials an additional linear or even superlinear increase of the conductivity was detected at higher frequencies and low temperatures [3-13]. In the present work we show data on the electrical properties of dipolar, ionically, and electronically conducting materials, measured over an extremely broad frequency range. We find that for all kinds of disordered matter in addition to the UDR a superlinear power law (SLPL) shows up, which extends well up to optical frequencies, before entering the regime of microscopic excitations. We conclude that in the full spectral range from audio to optical frequencies, *all* classes of disordered matter exhibit an astonishingly similar answer to electromagnetic fields, characterized by the succession of a sublinear and a superlinear power-law.

The response of matter to ac electrical fields is usually described in terms of the complex conductivity $\sigma^* = \sigma' + \sigma''$ or complex dielectric permittivity $\varepsilon^* = \varepsilon' - i\varepsilon''$. In the present letter, $\sigma'(\nu)$ and the dielectric loss $\varepsilon''(\nu)$ are considered, both related by $\sigma' = 2\pi\, \nu\, \varepsilon''\varepsilon_0$ ($\varepsilon_0$ the permittivity of free space). We performed measurements in an exceptionally broad frequency range by combining a variety of different experimental techniques, bridging the gap between classical dielectric and infrared spectroscopy (for details, see [12,14]). At low frequencies, electrode polarization plays a role and in the present work only data are shown where electrode contributions can be excluded [15].

Supercooled liquids, transferring into glasses at low temperatures, are classical examples of disordered matter. For dipolar supercooled liquids, recently the complete dielectric response up to infrared frequencies was determined [11,12]. As an example, in the upper part of Fig. 1 we show $\sigma'(\nu)$ for a series of temperatures in liquid and supercooled propylene carbonate (PC) [12]. The knees shifting through the frequency window correspond to peaks in $\varepsilon''(\nu) \sim \sigma'(\nu)/\nu$ (Fig. 2a). The position of these so-called $\alpha$-relaxation peaks characterizes the typical timescale of the motional dynamics of the dipolar molecules, closely related to the structural relaxation process, determining, e.g., the viscosity. Its continuous temperature shift demonstrates the slowing down of the molecular dynamics at the liquid-to-glass transition [12]. The UDR regime is revealed at the high-frequency side of the relaxation peaks, and in many supercooled liquids, including PC, can be identified with the so-called "excess wing" showing up as a second power law following the high-frequency flank of the $\alpha$-peak. It was recently demonstrated to be due to a second relaxation process of so-far unsettled origin [16], which in various other supercooled liquids leads to well-defined peaks at frequencies beyond the $\alpha$-peaks [17].

The lines through the curves of PC in Fig. 1 are fits with an empirical function describing the structural relaxation contribution [12], followed by the UDR, $\sigma' \sim \nu^s$ ($s < 1$). However, at high frequencies, fits and experimental curves deviate and a SLPL, $\sigma'(\nu) \sim \nu^n$, shows up as indicated by the dashed line, calculated with $n = 1.6$. The superlinear frequency dependence of $\sigma'(\nu)$ is best demonstrated plotting $\varepsilon''(\nu) \sim \sigma'(\nu)/\nu$, where the UDR and SLR regimes are expected to be separated by a well-defined minimum, which indeed is observed (Fig. 2a). Beyond the minimum, $\varepsilon''(\nu)$ continues increasing towards optical frequencies where a peak shows up, usually termed "boson peak", with a characteristic frequency typical for vibrational excitations. The overall dielectric response of supercooled liquids, described above, is also shared by the so-called plastic crystals, where the center of masses of the molecules reside on a translationally invariant lattice, but the molecules exhibit disorder with respect to their orientational degrees of freedom [18].

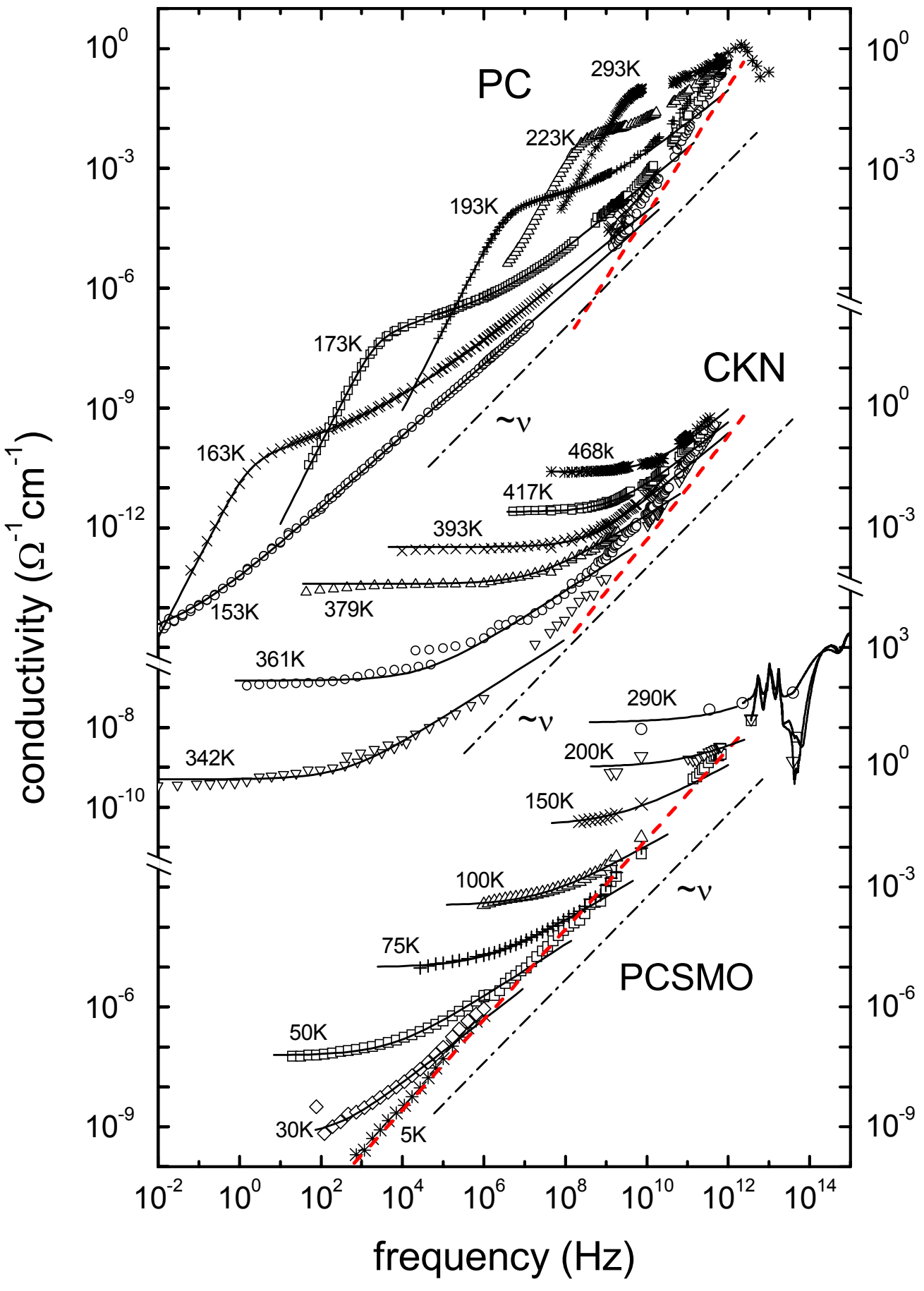

**Fig. 1.** Frequency-dependent conductivity at various temperatures for three different disordered materials. For the ionic conductor CKN and the electronic semiconductor PCSMO the solid lines below THz indicate fits with the sum of dc conductivity and the UDR. For the dipolar system PC the dc conductivity is replaced by a relaxation contribution. The dash-dotted lines demonstrate a slope of one. At high frequencies a transition from UDR to a SLPL, indicated by the dashed lines, is observed

In ionically conducting materials, the ac response is dominated by charge transport via hopping ions. As an example we show the behavior of glass forming $[Ca(NO_3)_2]_{0.4}[KNO_3]_{0.6}$ (CKN) [9] (Figs. 1 and 2b). In this case the low-frequency response is dominated by dc contributions, showing up as a low-frequency plateau in $\sigma'(\nu)$, respectively an initial $\nu^{-1}$ decrease in $\varepsilon''(\nu)$, which is followed by the UDR. For ionic conductors, the UDR is often ascribed to hopping conduction of the ionic charge carriers (see, e.g., [4,7,10]). The lines through the CKN curves in Fig. 1 are fits with $\sigma' = \sigma_{dc} + \sigma_0 \nu^s$. Again a SLPL of $\sigma'(\nu)$ towards the microscopic peak is clearly revealed (dashed line, $n = 1.3$), leading to a shallow minimum in $\varepsilon''(\nu)$ (Fig. 2b). A similar behavior was observed for other, not only glassy, but also crystalline, ionic conductors (see, e.g., [3-8,10]). In particular, in several experiments a high-frequency transition to a nearly linear increase of $\sigma'(\nu)$, corresponding to a so-called "nearly constant loss" was reported [3,8,10], which is also seen in CKN, e.g. at 342 K (Fig. 2b). However, measurements extending up to sufficiently high frequencies or low temperatures indeed reveal a SLPL [4,6,7], leading to a shallow minimum in $\varepsilon''(\nu)$.

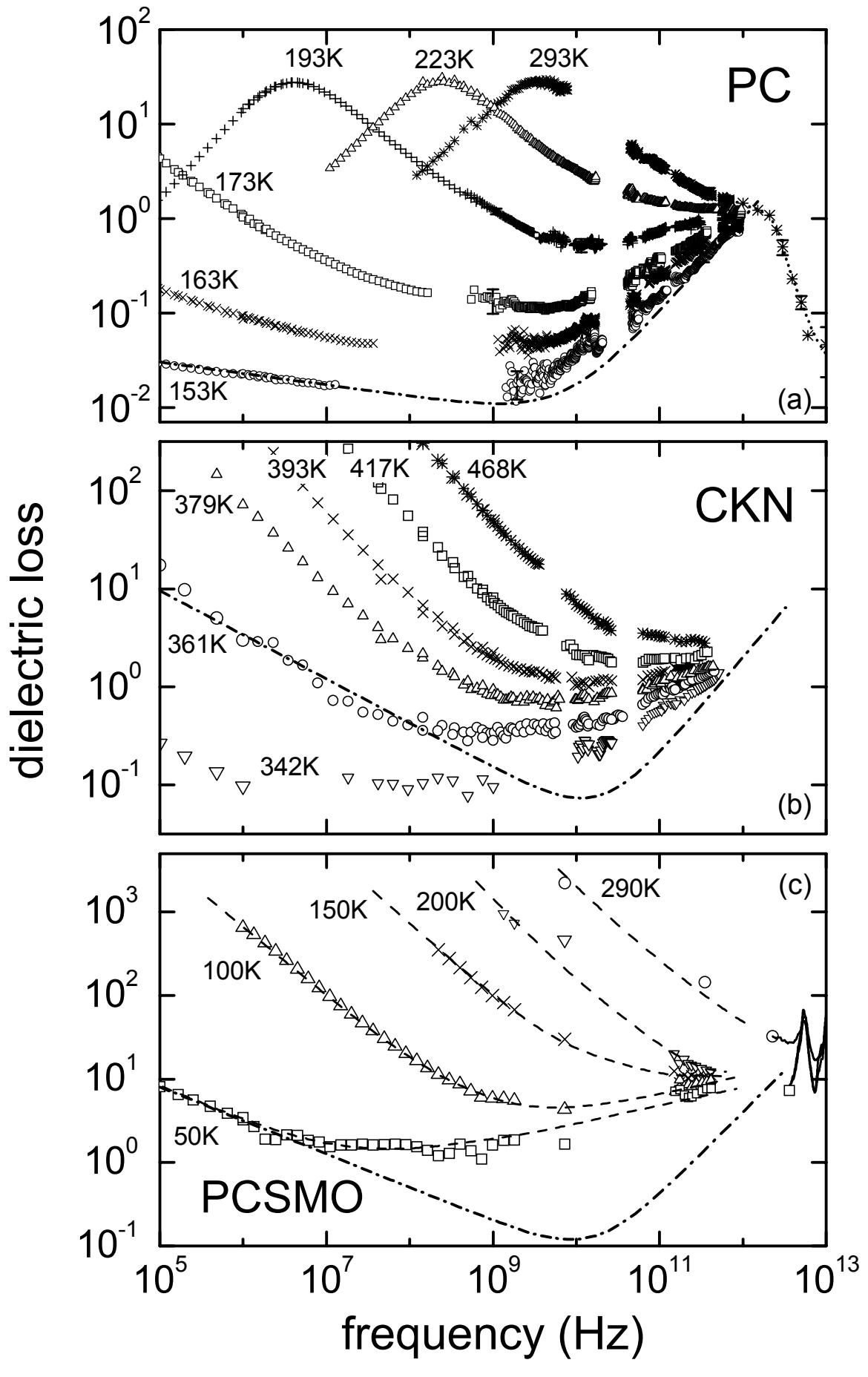

**Fig. 2.** Frequency-dependent dielectric loss for the same materials as in Fig. 1. The universal occurrence of a shallow loss minimum in disordered matter is demonstrated. The dashed lines are guides to the eyes. For PCSMO the results in the THz region are shown as solid lines. For PC and CKN the solid lines are fits of the minimum region with the MCT prediction.

Finally, and rather unexpected, we found that also electronic semiconductors, having substitutional disorder due to doping, show this typical succession of UDR and SLPL. We show results on a colossal magnetoresistance manganite, close to the metal-to-insulator boundary, namely $Pr_{0.65}(Ca_{0.8}Sr_{0.2})_{0.35}MnO_{0.35}$ (PCSMO) [19]. This material is especially suited due to its strong substitutional disorder. Again, at low frequencies we find a sequence of dc conductivity and UDR (solid lines in the lower part of Fig. 1). For doped and amorphous semiconductors, the UDR is usually ascribed to hopping conduction of Anderson-localized charge carriers [20] and theoretically explained, e.g., in the framework of Motts Variable-Range-Hopping model [21]. But for high frequencies, very similar to the behavior of supercooled liquids and ionic conductors, $\sigma'(\nu)$ smoothly passes into a SLPL extending towards THz frequencies (dashed line, $n = 1.13$). In $\varepsilon''(\nu)$ a shallow minimum shows up (Fig. 2c), before in the optical region phonon modes lead to sharp resonance-like features.

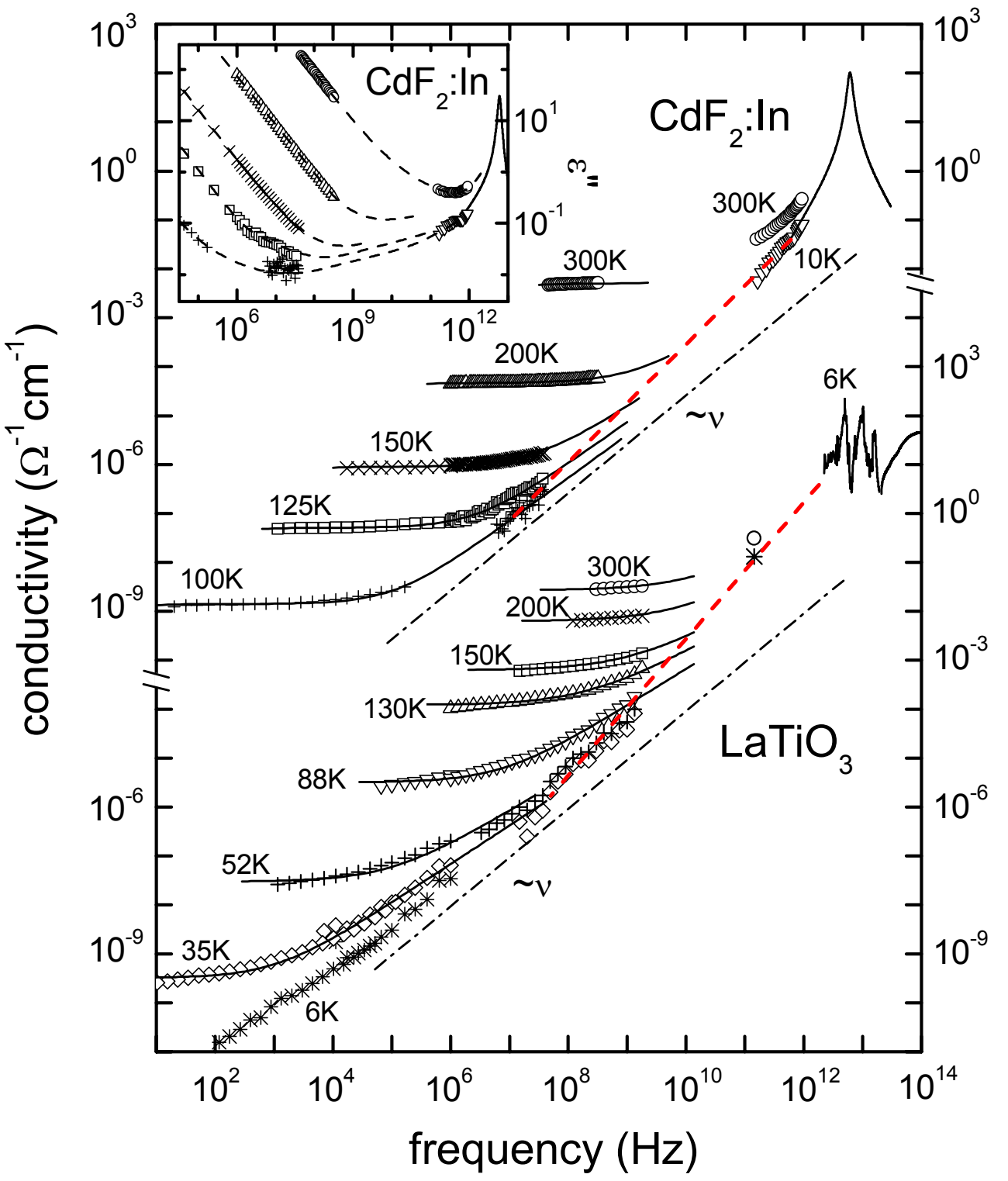

**Fig. 3.** Frequency-dependent conductivity at various temperatures for semiconducting $CdF_2$:In [23] and $LaTiO_3$. The results in the THz region are shown as solid lines. The solid lines below THz indicate fits with the sum of dc conductivity and the UDR. The dash-dotted lines demonstrate a slope of one; the dashed lines indicate the SLPL. The inset shows the high-frequency dielectric loss minimum of $CdF_2$:In; the dashed lines are guides to the eyes.

One of the very few examples of measurements in a similarly broad frequency range on electronic conductors is given in [22], where quite similar results were obtained for a polaronic semiconductor. In particular, an increase of $\sigma'(\nu)$ steeper than linear was observed for high frequencies, which also corresponds to a shallow minimum in $\varepsilon''(\nu)$. Further examples of electronic conductors are given in Fig. 3. An interpolation between low- and high-frequency results reveals that for $CdF_2$:In [23] a SLPL most likely is present, too (dashed line, $n = 1.2$). The inset shows, that indeed a shallow minimum of $\varepsilon''(\nu)$ is observed. $CdF_2$:In is a doped semiconductor with shallow donor states. The results on the pure single-crystalline Mott-Hubbard insulator $LaTiO_3$ demonstrate that the UDR and SLPL (dashed line, $n = 1.5$) also can be observed in systems with marginal disorder only. In this case a slight off-stoichiometry of the oxygen content or impurities at a ppm level are sufficient to produce the observed typical response, common to all disordered matter.

The unified view that can be condensed out of Figs. 1-3 is presented in Fig. 4, where we schematically show $\sigma'(\nu)$ and $\varepsilon''(\nu)$ for disordered matter. At the lowest frequencies, for materials with reorienting dipoles relaxational behavior shows up, while dc conductivity is observed for materials with free charge carriers (e.g. ions, electrons, or holes). At higher frequencies, all classes of disordered matter show a succession of a sublinear power-law at intermediate and an only weakly temperature-dependent SLPL at high frequencies, with the crossover frequency shifting to higher frequencies on increasing temperature. This behavior, indicated by the solid lines, can even better be identified by plotting the dielectric loss, where these two regimes are separated by a shallow minimum (Fig. 4b).

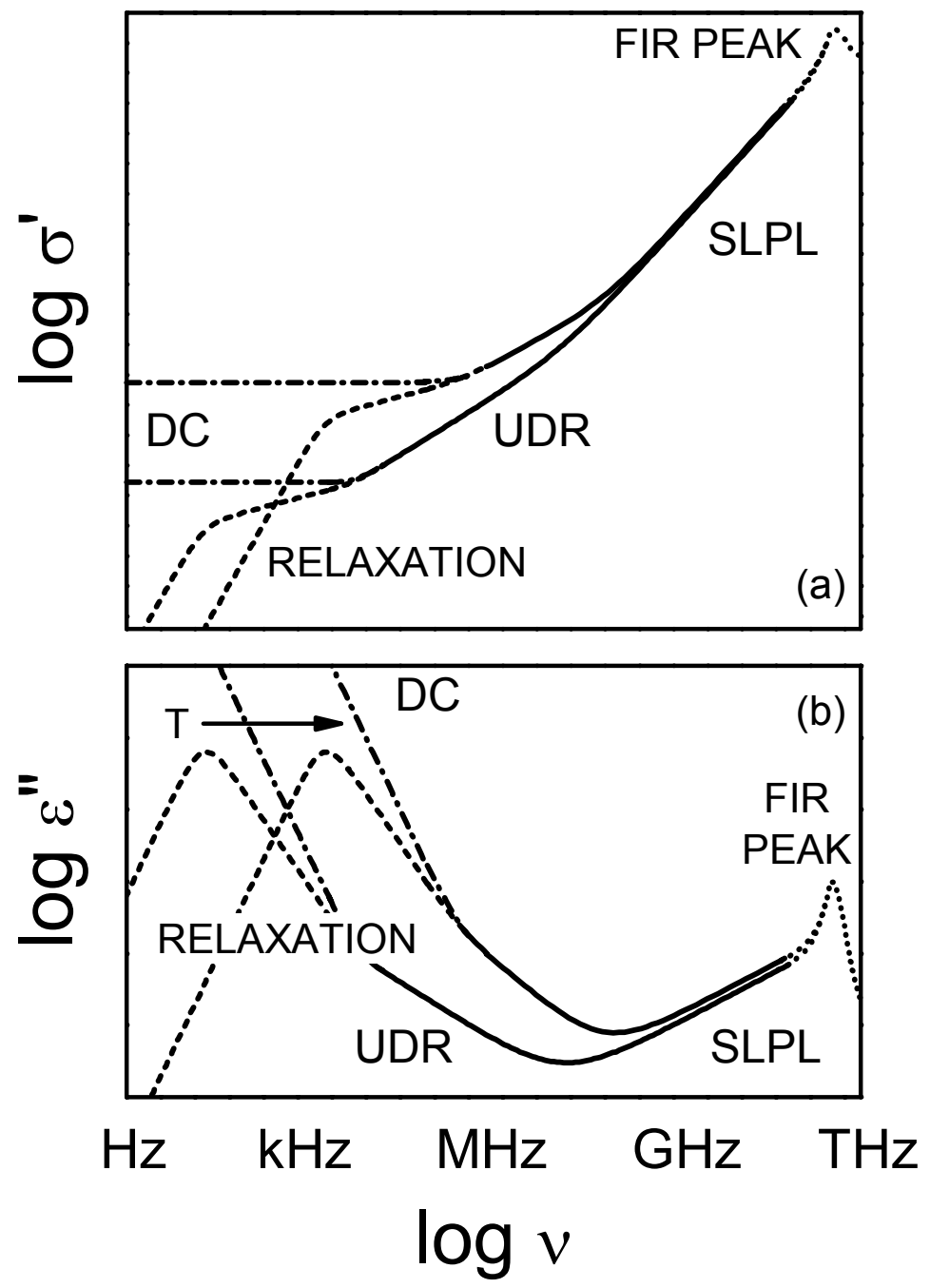

**Fig. 4.** Schematic view of the frequency-dependent response of disordered matter in double-logarithmic representation. Two curves for two different temperatures are shown. At low frequencies for non-conducting dipolar systems relaxational behavior is observed (dashed lines; here the case of a system with excess wing is shown), which is replaced by dc conductivity for conducting systems (dash-dotted lines). The region of universality is indicated by the solid lines. In the far-infrared (FIR) region, microscopic excitations (phonon resonances or boson peak, dotted line) show up. At the abscissa a rough indication of the frequency ranges is given.

Concerning the origin of the observed SLPL we want to emphasize that it cannot be explained by a simple transition from nearly linear UDR behavior to a quadratic or stronger frequency dependence, which can be assumed for the low-frequency flank of the microscopic excitations. This is demonstrated by the dash-dotted lines in Fig. 2, which were calculated by adding the UDR contribution and a linear increase of $\varepsilon''$ (corresponding to $\sigma' \sim \nu^2$). Clearly the experimental data cannot be described in this way and there must be an additional contribution – the SLPL. Only in PC at the lowest temperature, where $n$ is largest, the experimental data are nearly matched by this ansatz. However, one has to be aware that a $\varepsilon'' \sim \nu$ increase corresponds to a Debye relaxational behavior and it is reasonable to assume that the actual increase towards the boson peak is steeper. In literature a number of explanations for a SLPL can be found: For supercooled liquids, the mode-coupling theory predicts a shallow high-frequency minimum in $\varepsilon''(\nu)$ with a high-frequency flank following a power law $\nu^a$, with $a < 0.4$ [24]. An alternative explanation for the shallow minimum was proposed in terms of a ubiquitous nearly constant loss contribution (comprising also a superlinear $\sigma'(\nu)$ with $1.2 < n < 1.3$), which was ascribed to a vibrational relaxation process [25]. Finally, some theories on hopping conduction may also allow for a superlinear behavior in certain cases [20,26]. It should be noted that the observed SLPL most likely is not connected to the photon-assisted hopping, recently observed in various semiconductors at low temperatures and high frequencies and characterized by an approximately linear or even quadratic frequency dependence of $\sigma'(\nu)$ [13]. A transition from phonon- to photon-assisted hopping should occur for $h\nu > k_B T$, a condition that clearly is not fulfilled for most of the regions where the SLPL is observed in the present work. Overall, none of the above explanations for SLPL can account for its occurrence in all the different types of materials reported here. In view of the similarity of the ac response of doped semiconductors and systems close to a metal-to-insulator transition to that of canonical glass formers, it seems relevant to further investigate glassy behavior of electrons. And indeed, recently it has been shown that glassy behavior of electrons emerges before the charge carriers localize, and in addition, that Anderson localization enhances the stability of the glass phase, while Mott localization tends to suppress it [27]. It certainly would be an attractive idea to apply concepts, successfully used for canonical glass-formers (as e.g. the MCT), to disordered electronic systems, but as these theories consider mainly density-density correlations further theoretical work seems necessary.

In conclusion, by measuring the extreme broadband dielectric response of different disordered materials, we found an astonishing universality of their response to electromagnetic fields extending over a broad dynamic range up to the infrared regime. Jonscher´s UDR, characterized by a sublinear power law at intermediate frequencies, is followed by a superlinear power law with an exponent significantly smaller than two. This universality is valid for liquids, supercooled liquids, glasses, ionic melts, glassy and crystalline ionic conductors, orientationally disordered crystals, and doped semiconductors, with the electromagnetic field acting on diffusing or reorienting dipoles, hopping ions, and localized holes or electrons. Thus disordered condensed matter seems to exhibit a universal response to ac electrical fields that is far more general than thought until now. We feel that this common behavior is too striking to be accidental and believe that one underlying principle governs the response of disordered matter to electromagnetic fields.

The authors gratefully acknowledge stimulating discussions with K. L. Ngai, A. V. Pronin, A. I. Ritus, and A. A. Volkov. This work was partly supported by the BMBF via VDI/EKM 13N6917A and by the Deutsche Forschungsgemeinschaft via the SFB 484.